\documentclass[conference]{IEEEtran}
\IEEEoverridecommandlockouts
\usepackage{cite}
\usepackage{amsmath,amssymb,amsfonts}
\usepackage{algorithm}
\usepackage{algpseudocode}
\usepackage{subcaption}
\usepackage{graphicx}
\usepackage{textcomp}
\usepackage{xcolor}
\usepackage{siunitx}
\usepackage{microtype}
\usepackage[hidelinks]{hyperref}
\usepackage{tikz}
\usepackage{booktabs}
\input{mysymbol.sty}
\def\BibTeX{{\rm B\kern-.05em{\sc i\kern-.025em b}\kern-.08em
    T\kern-.1667em\lower.7ex\hbox{E}\kern-.125emX}}

\begin{document}


\title{Sampling with Shielded Langevin Monte Carlo Using Navigation Potentials\\
}
\vspace{-0.6em}
\author{%
  \IEEEauthorblockN{Nicolas Zilberstein$^{\dagger}$, Santiago Segarra$^{\dagger}$, Luiz Chamon$^{\star}$}
  \IEEEauthorblockA{$^{\dagger}$Rice University, USA \hspace{4cm}
                    $^{\star}$\'{E}cole polytechnique, Institut Polytechnique de Paris, FR}
\thanks{This research was partially sponsored by ARO (W911NF-17-S-0002) and NSF (CCF-2340481). The views and conclusions contained in this document are those of the authors and should not be interpreted as representing the official policies, either expressed or
implied, of the Army Research Office, the U.S. Army, or the U.S. Government. The U.S. Government is authorized to reproduce and distribute reprints for Government purposes, notwithstanding any copyright notation herein.}
}

\IEEEaftertitletext{\vspace{-1\baselineskip}}

\maketitle

\begin{abstract}
We introduce shielded Langevin Monte Carlo (LMC), a constrained sampler inspired by navigation functions, capable of sampling from unnormalized target distributions defined over punctured supports. In other words, this approach samples from non-convex spaces defined as convex sets with convex holes. This defines a novel and challenging problem in constrained sampling. To do so, the sampler incorporates a combination of a spatially adaptive temperature and a repulsive drift to ensure that samples remain within the feasible region. Experiments on a 2D Gaussian mixture and multiple-input multiple-output (MIMO) symbol detection showcase the advantages of the proposed shielded LMC in contrast to unconstrained cases.
\end{abstract}


\section{Introduction}
\vspace{-0.1cm}
Sampling plays a central role in statistics, supporting tasks such as estimation, decision making, and uncertainty quantification~\cite{van2021bayesian, ermon2019}.
In modern machine learning (ML), it is crucial, forming the backbone of generative models~\cite{song2021scorebased} and probabilistic inference methods~\cite{vargasdenoising}.
In many real-world settings, however, the target distribution is subject to constraints, be they physical (e.g., feasibility or system dynamics) or user-specified (e.g., safety requirements~\cite{khatib1986real, paternain2022safe}).
Examples include motion planning in robotics~\cite{janner2022planning}, molecular structure prediction~\cite{schneuing2024structure}, and image generation under restricted domains, such as avoiding violent content~\cite{kim2025training}.
The goal of this work is to design a sampling algorithm that guarantees that the generated samples satisfy such underlying constraints.

Markov chain Monte Carlo (MCMC) methods~\cite{robert1999monte} form a well-established class of techniques for sampling from complex distributions known only up to a normalizing factor.
Among them, one of the most widely used is Langevin Monte Carlo (LMC)~\cite{Roberts1996ExponentialCO, dalalyan2019}, an iterative method that samples from a target distribution by exploiting access to the score function (i.e., the gradient of the log-density).
However, classical sampling algorithms such as LMC do not naturally incorporate constraints on the generated samples: they are not designed to enforce structural restrictions or to handle safety requirements on the target distribution.
To address this limitation, several extensions have been proposed, based on projection schemes~\cite{bubeck2018sampling}, mirror/proximal maps ~\cite{hsieh2018mirrored, zhang2020wasserstein,salim2020primal}, barriers~\cite{kook2022sampling, noble2023unbiased}, penalties~\cite{gurbuzbalaban2024penalized}, and primal-dual methods~\cite{chamon2024constrained}.
Though effective, these methods are primarily geared towards convex constraints, particularly those tackling support constraints. 
Indeed, even evaluating projections, barriers, and proximal maps is infeasible in non-convex settings.

In this work, we propose a Langevin-based algorithm designed specifically for \textit{distributions with non-convex support}.
Explicitly, for distributions defined over convex domains containing convex holes (see Fig.~\ref{fig:scheme}).
Drawing inspiration from control theory~\cite{paternain2017navigation}, we build upon the classical Rimon–Koditschek (RK) artificial potential~\cite{koditschek1990robot}, originally developed for navigation in spaces with convex obstacles.
Incorporating this potential leads to two key modifications of the Langevin diffusion.
First, it modulates the diffusion term by incorporating a spatially varying temperature that depends on the distance to obstacles or holes.
This adaptive temperature reduces randomness near puncture boundaries.
Second, it includes an additional drift term that induces a repulsive force around forbidden regions, ensuring that the sampling trajectories remain within the feasible set.
These components together define the \emph{shielded Langevin Monte Carlo} sampler, a principled approach for sampling from distributions supported on punctured domains.

\vspace{1mm}
\noindent {\bf Contributions.}
Our main contributions are twofold:\\
1) We introduce \emph{shielded LMC}, a Langevin-based constrained sampler that enables sampling from non-convex punctured spaces by combining the RK potential with Langevin dynamics.\\
2) We illustrate the use of the proposed sampler on two settings: $(i)$ a 2D Gaussian mixture model and $(ii)$ symbol detection in multiple-input multiple-output (MIMO) systems, validating that the samples satisfy the constraints.




\section{Problem formulation and background}
\label{sec:problem}

\begin{figure}[t]
\centering
\begin{minipage}{0.3\textwidth}
	\centering
	\resizebox{2.2cm}{!}{\begin{tikzpicture}
			\useasboundingbox (-4.5, -5) rectangle (4.5, 6.2);
			
			\path [fill=gray!30, draw=black, even odd rule]
			(-4.5, -4.5) rectangle (4.5, 4.5)
			
			(-2, -1.5) circle (1.2)
			
			(2.5, 2) ellipse (1 and 1.5)
			
			(1.5, -2.5) rectangle (3.5, -1.5);
			
			\node at (-2, -1.5) {\Huge$\ccalO_1$};
			\node at (2.5, 2) {\Huge$\ccalO_2$};
			\node at (2.5, -2) {\Huge$\ccalO_3$};
			
			\node at (-2, -3.5) {\Huge $\ccalC = \cup \ccalO_i$};
			
			\node at (-2.2, 3) {\Huge $\ccalF = \reals^d \setminus \ccalC$};
			
			\node at (0, 5) {\Huge $\pi_{\text{constr}}(\bbx) \propto p(\bbx)\mathbb{I}\{\bbx \in \ccalF\}$};
	\end{tikzpicture}}
    \hspace{0.2cm}
	\resizebox{2.2cm}{!}{\begin{tikzpicture}
			\useasboundingbox (-4.5, -5) rectangle (4.5, 6.2);
			
			\draw [black!50, dashed, thin] (-4.5, -4.5) rectangle (4.5, 4.5);
			\node at (-3.5, 4) {\Huge $\reals^d$};
			
			\draw [fill=gray!30, draw=black] (0, 0) ellipse (3.5 and 3);
			
			\node at (0, 0) {\Huge $\ccalC$};
			
			\node at (0, 5) {\Huge $\pi_{\text{constr}}(\bbx) \propto p(\bbx)\mathbb{I}\{\bbx \in \ccalC\}$};
	\end{tikzpicture}}
\end{minipage}
\caption{\footnotesize{Comparison of constraint types. Left: Our setting, where convex obstacles carve out forbidden regions, yielding a non-convex feasible space from which we must sample. Right: The classical case, where sampling is restricted to a convex feasible region within the ambient space.}}
\vspace{-0.4cm}
\label{fig:scheme}
\end{figure}
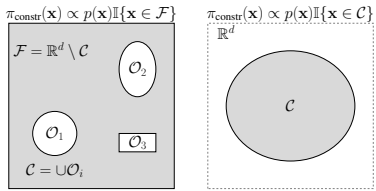

We consider the problem of sampling from a punctured domain.
Specifically, let $p(\bbx)$ be a target (unconstrained) distribution supported on a space $\ccalX$.
Typically, $p(\bbx)$ is known only up to a normalization factor, i.e., we know only $U(\bbx)$ s.t. $p(\bbx) \propto e^{-U(\bbx)}$.
Our goal is to obtain samples from the \textit{constrained} distribution 
\begin{equation}
    \label{eq:target_distr}
    p_{\ccalF}(\bbx) \propto p(\bbx)\, \mathbb{I}\{\bbx \notin \ccalC\},
\end{equation}
where $\ccalC = \bigcup_{i=1}^n \ccalO_i$ denotes a union of non-overlapping \textit{obstacles} $\ccalO_i$ we seek to avoid. 
Each obstacle $\ccalO_i$ is assumed to be \textit{closed} and \textit{convex} and we associate with it a function $\beta_i(\bbx)$ such that
\begin{equation}
\label{eq:obstacle}
    \ccalO_i = \{\bbx \in \ccalX : \beta_i(\bbx) < 0\}.
\end{equation}
In addition, we define the aggregated function $\beta(\bbx) = \prod_{i=1}^n \beta_i(\bbx)$ that characterizes the overall constraint region $\ccalC$.
Hence, our goal is to sample from $p_{\ccalF}$, whose support corresponds to the \textit{punctured space}, or \textit{free space} $\ccalF = \ccalX \setminus \ccalC$.
In Fig.~\ref{fig:scheme}, we provide an illustrative $\ccalF$ and the corresponding obstacles $\ccalO_i$.
Note that the support $\ccalF$ of $p_{\ccalF}$ is non-convex despite the convexity of $\ccalC$.
Indeed, contrary to the traditional constrained sampling problem, we aim to sample from $p_{\ccalF}$ in~\eqref{eq:target_distr} (Fig.~\ref{fig:scheme}, left) as opposed to $p(\bbx) \, \mathbb{I}\{\bbx \in \ccalC\}$ (Fig.~\ref{fig:scheme}, right). 
In what follows, we introduce the two main ingredients that form the basis of our proposed constrained sampling method.

\subsection{Langevin diffusion}
\label{subsec:langevin}

The Langevin diffusion is the stochastic process on $\mathbb{R}^d$ that satisfies the stochastic differential equation
\begin{align}\label{eq:ct_langevin}
    \text{d}\bbx_t &= -\nabla U(\bbx_t) + \sqrt{2\tau}\text{d}\bbW_t,
\end{align}
where $\bbW_t$ is a standard $d$-dimensional Brownian motion, $U \in \ccalC^2(\mathbb{R}^d)$ is called the \textit{potential}, and $\tau$ is a temperature parameter.
The dynamic in~\eqref{eq:ct_langevin} describes the path of a particle moving according to Newton's second law under a force $\nabla U(\bbx_t)$ while in contact with a thermal reservoir, i.e., collisions with other particles, represented by the random term (Brownian motion).
Under mild conditions, it can be shown that the invariant distribution $\pi(\bbx)$ of this continuous-time process is proportional to $\exp (-\tau^{-1}U(\bbx))$~\cite{pavliotis_book}.
In particular, given a target distribution $p(\bbx)$ and fixing $\tau = 1$, we can obtain an MCMC sampler by letting $U(\bbx) = -\log p(\bbx)$.
The celebrated MCMC algorithm \cite{robert1999monte, Roberts1996ExponentialCO} known as unadjusted Langevin algorithm, is obtained as the Euler-Maruyama discretization of~\eqref{eq:ct_langevin} and is described by the discrete-time equation
\begin{equation}\label{eq:langevin}
	\bbx_{k+1} = \bbx_k + \epsilon_k \nabla_{\bbx_k}\log p(\bbx_k) + \sqrt{2\tau \epsilon_k}\, \bbw_k,
\end{equation}
where $k$ is an iteration index, $\epsilon_k > 0$ is the discretization step size (which can be time-varying), and $\bbw_k\sim\ccalN(\pmb{0}, \bbI_d)$.
Under some regularity conditions, the distribution of $\bbx_k$ converges to $\pi(\bbx) \propto p(\bbx)^{1/\tau}$ as $\epsilon \rightarrow 0$ and $k \rightarrow \infty$.
Although this result is asymptotic, non-asymptotic convergence results have been obtained under additional conditions on the target distribution~\cite{dalalyan2019}.

\subsection{Rimon-Koditschek (RK) navigation potential} \label{subsec:navigation}

The problem of navigation in the presence of obstacles is fundamental in robotics and control theory. 
The goal is to design trajectories that guide an agent (e.g., a robot or drone) from an initial configuration in state-space to a desired goal configuration, while avoiding forbidden regions of the configuration space (i.e., obstacles). 
In the absence of obstacles, this problem can be described in terms of a (convex) potential $U(\bbx)$ whose minimum $U(\bbx^*) = 0$ describes the goal configuration.
Then, its solution can be obtained by implementing the gradient flow
$\dot{\bbx}_t = -\nabla_{\bbx_t}U(\bbx_t)$, where the minimum $\bbx^*$ corresponds to the goal configuration.

In the presence of obstacles $\ccalO_i$, which is the setting of this paper, the challenge is to design a potential that both attracts the agent to the goal and repels it from obstacles.
A typical approach is to construct a navigation function which entails an \textit{artificial potential} that achieves both aforementioned desiderata.
It can be shown that implementing gradient flow along the navigation function indeed achieves the goal while avoiding obstacles as long as it satisfies the following properties~\cite{koditschek1990robot}.
\vspace{2mm}
\begin{definition}[Navigation Function]
A function $\phi: \ccalX \to [0,1]$ is a navigation function on the configuration space $\ccalX$ if it satisfies:
\begin{enumerate}
    \item $\phi$ is smooth (at least $C^2$) on $\ccalX$
    \item $\phi$ has a unique minimum at the goal configuration
    \item $\phi$ increases to 1 on the boundary of obstacles
    \item The only critical point in the interior of the free space is $\bbx^*$.
\end{enumerate}
\end{definition}
\vspace{2mm}
For instance, the RK artificial potential from~\cite{koditschek1990robot} is a proper navigation function. 
Explicitly, it is given by
\begin{equation}\label{eq:rk_potential}
    \phi_\alpha(\bbx) = \frac{U(\bbx)}{\left(U(\bbx)^\alpha + \beta(\bbx)\right)^{1/\alpha}},
\end{equation}
where $\alpha > 0$ is a tuning parameter and $\beta$ is defined as in~\eqref{eq:obstacle}.
The RK potential combines a given potential $U(\bbx)$, which encodes the configuration goal, with a repulsive potential $\beta(\bbx)$, encoding the obstacle information. 
Therefore, the goal $\bbx^*$ can be reached w/o hitting obstacles $\ccalO_i$ by implementing $\frac{\text{d} \bbx_t}{\text{d} t} = -\nabla_{\bbx_t}\phi_{\alpha}(\bbx_t)$, where the gradient is given by
\begin{equation}\label{eq:rk_gradient}
    \nabla_{\bbx}\phi_\alpha(\bbx) = \frac{\beta(\bbx) \nabla_{\bbx}U(\bbx) - \frac{U(\bbx)}{\alpha}\nabla_{\bbx}\beta(\bbx)}{\left(U(\bbx)^\alpha + \beta(\bbx)\right)^{1 + 1/\alpha}}.
\end{equation}
This gradient has an intuitive interpretation: 
\begin{itemize}
    \item Near the goal $x^*$ (where $U(\bbx) \approx 0$), the original gradient $\nabla_{\bbx}U(\bbx)$ dominates pulling the agent toward the goal;
    \item Near an obstacle (where $\beta(\bbx) \approx 0$), the repulsive term $-\nabla_{\bbx}\beta(\bbx)$ dominates pushing the agent away from the obstacle;
    \item The tuning parameter $\alpha$ controls the balance between attraction and repulsion.
\end{itemize}


We will leverage this approach from navigation to build shielded LMC, a constrained sampler for spaces with convex obstacles.

\section{Shielded LMC sampler}

Recall that our goal is to sample from $p_{\ccalF}(\bbx)$ in~\eqref{eq:target_distr} supported on the free space $\ccalF$. 
To do this, we will use the Langevin dynamics introduced in Section~\ref{subsec:langevin}. 
However, we need to incorporate the constraint $\mathbb{I}\{\bbx\notin \ccalC\}$, which is non-differentiable and, therefore, cannot be directly plugged into the dynamic.
To overcome this challenge, we leverage the artificial potential framework from Section~\ref{subsec:navigation} to construct a constrained variant of the Langevin dynamics.
Hence, from now on, we consider $U(\bbx) = -\log p(\bbx) + C$ in~\eqref{eq:rk_potential}, where we assume that $p(\bbx)$ is log-concave (although in the numerical experiments we will consider cases where this does not hold) and $C$ is such that $U(\bbx) > 0 \;\forall\; \bbx$.

A natural first attempt would be to replace the unconstrained distribution in~\eqref{eq:langevin} with the RK potential $\phi_\alpha$ from~\eqref{eq:rk_potential}.
This yields the modified dynamics
\begin{align}\label{eq:ct_rk_wrong_langevin}
\text{d}\bbx_t = -\nabla \phi_\alpha(\bbx_t)\text{d}t + \sqrt{2\tau}\text{d}\bbW_t.
\end{align}
While this modification may appear reasonable, it is in fact incorrect due to the isotropic nature of the noise. Specifically, as the trajectory approaches an obstacle, the noise component remains isotropic and of constant variance, which can still push the particle into the forbidden region. Hence, a more careful treatment of the stochastic term is required to ensure the dynamic generates trajectories that lie in the free space $\ccalF$.
We tackle this next.
\subsection{Building the sampler by using time-varying noise}
\label{subsec:rk_sampler}

In order to control the diffusion term, we modify not only the potential but also the variance of the noise. 
In particular, we propose the dynamics
\begin{align}\label{eq:rk_langevin_cts}
\text{d}\bbx_t = -\nabla \phi_\alpha(\bbx_t)\text{d}t + \sqrt{2\tau\beta^2(\bbx_t)}\text{d}\bbW_t,
\end{align}
where $\beta(\bbx_t)$ is the aggregated function encoding the obstacles~(see Section~\ref{sec:problem}).
We can interpret this new variance as a \textit{spatially-aware} temperature function that modulates the diffusion term based on the proximity to obstacles. 
After discretization, we obtain the update rule
\begin{align}\label{eq:rk_langevin_disc}
\bbx_{k+1} \! = \! \bbx_k \!  + \! \eta_k&\left[\beta(\bbx_k)\nabla_{\bbx_k} \log p(\bbx_k)\!  - \! \frac{\log p(\bbx_k) \nabla_{\bbx_k}\beta(\bbx_k)}{\alpha}\right]
\\  \nonumber
& + \sqrt{2\eta_k\tau\beta^2(\bbx_k)}\bbw_k,
\end{align}
where $\bbw_k \sim \mathcal{N}(\boldsymbol{0}, \bbI)$.
Note that we incorporated the strictly positive denominator in~\eqref{eq:rk_gradient} into the step size $\eta_k$.
We refer to this modified dynamics as the \emph{shielded LMC sampler}.
The value of the parameter $\alpha > 0$ controls the behavior close to the optimum (i.e., the bias of the algorithm), trading off between the pure Langevin dynamics~\eqref{eq:langevin} and the repulsive strength of the obstacle.
In particular, if $\alpha$ is too small, the repulsion acts over a large neighborhood around the obstacle and can slow down the convergence to the stationary distribution.
On the other hand, if $\alpha$ is too large, the repulsion is very sharp (and only affects a narrow region near the obstacle), which can cause samples to get stuck at the boundaries of the obstacle.
We illustrate this compromise in the numerical experiments (Section~\ref{sec:experiments}).

Once $\alpha$ is fixed, we can analyze the behavior of the shielded LMC sampler in the free space~$\ccalF$.
To do this, we look at how the potential and spatially-aware temperature interact across different regions:

\begin{figure*}[t]
    \centering
	\begin{subfigure}{.3\textwidth}
    	\centering
    	\includegraphics[width=0.65\textwidth]{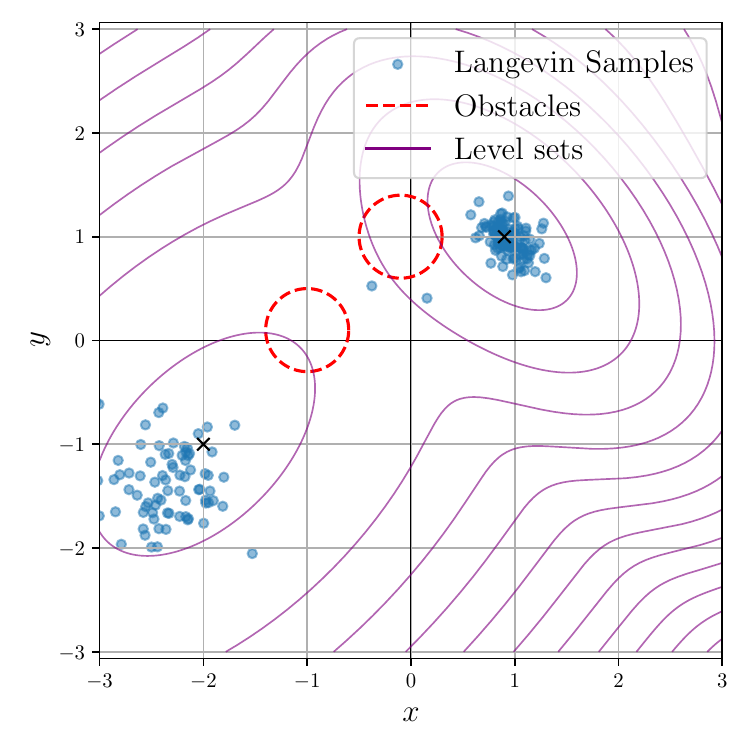}
    	\vspace{-0.15in}
    	\caption{$\alpha = 0.1$}
    	\label{fig:NMSE-discretization_methods}
	\end{subfigure}%
	\centering
	\begin{subfigure}{.3\textwidth}
    	\centering
    	\includegraphics[width=0.65\textwidth]{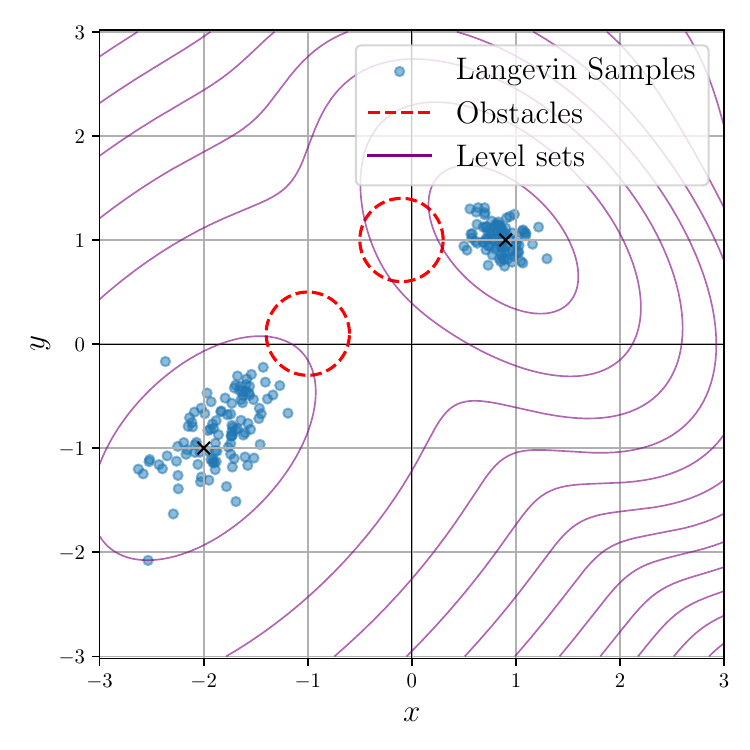}
    	\vspace{-0.15in}
    	\caption{$\alpha = 1$}
    	\label{fig:NMSE-langevin-order}
	\end{subfigure}
	\centering
	\begin{subfigure}{.3\textwidth}
    	\centering
    	\includegraphics[width=0.65\textwidth]{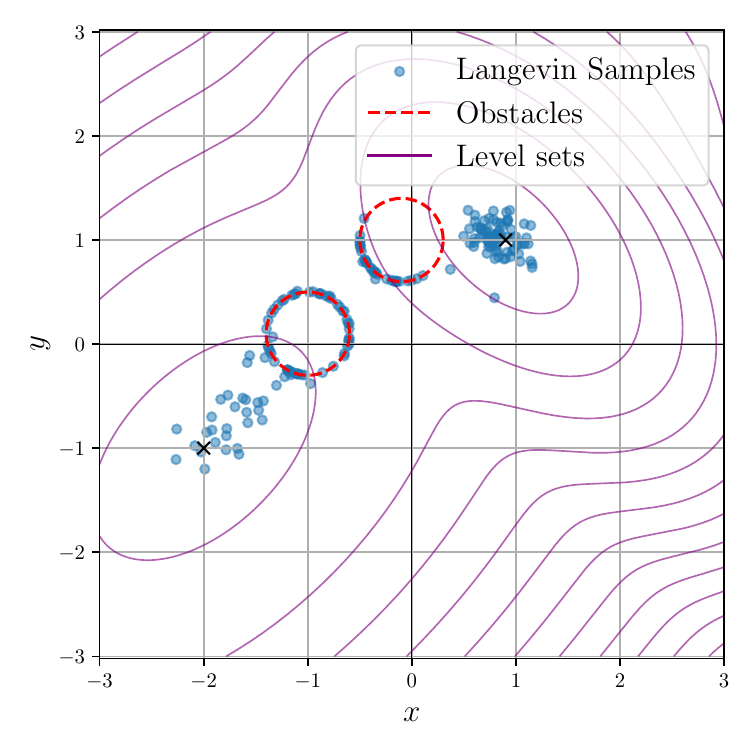}
    	\vspace{-0.15in}
    	\caption{$\alpha = 7$}
    	\label{fig:Hestimation_methods}
	\end{subfigure}
	\vspace{-0.02in}
	\caption{ {\small Impact of parameter $\alpha$ on shielded LMC with obstacles. Blue dots are Langevin samples after 50k iterations, red dashed lines are obstacles, and black crosses mark mode means; contours show the target distribution.(a) $\alpha = 0.1$: repulsion is too strong, pushing samples away from true modes. (b) $\alpha = 1$: good balance between target sampling and obstacle avoidance. (c) $\alpha = 7$: repulsion is too weak, causing samples to collapse onto obstacle boundaries where gradients vanish.}}
	\vspace{-0.2in}
	\label{figs_gaussian_case}
\end{figure*}

\begin{enumerate}
    \item In the free space far from both the goal and the obstacle, the term $\beta(\bbx)\nabla \log p(\bbx)$ dominates. 
    Since $\beta(\bbx) > 0$, the particle moves according to~\eqref{eq:langevin} with step size $\eta \beta(\bbx)$.
    As a result, the sampler distributes the particle according to $p(\bbx)$ while remaining largely unaffected by obstacles.

    \item Near obstacle boundaries, where $\beta(\bbx) \to 0^+$, the attractive term $\beta(\bbx)\nabla \log p(\bbx)$ disappears and the gradient becomes dominated by the repulsive component $-\log p(\bbx) \nabla \beta(\bbx)$. 
    This induces a strong repulsion that prevents samples from crossing into forbidden regions.

\end{enumerate}

\setlength{\textfloatsep}{2pt}
\begin{algorithm}[t]
    \caption{Shielded LMC}\label{alg:rk_sampler}
    \scalebox{0.8}{
    \begin{minipage}{\linewidth}
    \begin{algorithmic}[1]
        \Require $\nabla\log p(x), \beta(x), \tau$; Optional if available $\log p(x)$%
        \State Initialize $\bbx_0$ at random
        \For{$k \leftarrow 1\; \text{to}\;  K$}
            \State Compute $\beta(\bbx_k), \nabla_{\bbx_k} \beta(\bbx_k)$
            \State Compute $\nabla_{\bbx_k}\log p(\bbx_k)$ {\small{(If available, compute $\log p(\bbx_k)$)}}
            \State Draw $\bbw_k \sim \ccalN(0, \bbI)$
            \State Compute 
            \begin{align*}\bbx_{k+1} = \bbx_k &+ \eta_k [\beta(\bbx_k)\nabla_{\bbx_k} \log p(\bbx_k) - \frac{\log p(\bbx_k) \nabla_{\bbx_k}\beta(\bbx_k)}{\alpha}]\\ &
 + \sqrt{2\eta_k\tau\beta^2(\bbx_k)}\bbw_k
            \end{align*}
            \EndFor \\
        \Return $\bbx_K$
    \end{algorithmic}
    \end{minipage}
    }
\end{algorithm}
Overall, these two regimes allow the sampler to adapt to different regions of the space, balancing broad exploration when far from obstacles with stronger constraint avoidance as it approaches them.
The final algorithm is shown in Alg.~\ref{alg:rk_sampler}.


\vspace{2mm}
\noindent \textbf{Dependency on $\log p(\bbx)$.}
An alternative interpretation of the update in~\eqref{eq:rk_langevin_disc} is as a classical Langevin dynamic like~\eqref{eq:langevin} with an effective step size $\eta_k \beta(\bbx_k)$, augmented by an additional drift term
$-\frac{\log p(\bbx_k)\,\nabla_{\bbx_k}\beta(\bbx_k)}{\alpha}$.
This extra term explicitly depends on the target density $p(\bbx)$, which is generally intractable and, even when known, is available only up to a normalization constant.
In such cases, a practical workaround is to approximate $\frac{\log p(\bbx)}{\alpha}$ by a global constant $\bar{\alpha}$ that is independent of $\bbx$, thereby avoiding the need to evaluate $\log p(\bbx)$ during sampling.

\section{Numerical results} \label{sec:experiments}

We describe two experiments that demonstrate the advantages of our proposed sampler and how it compares with the unconstrained case.
First, in Section~\ref{sub:gmm} we analyze its behavior when sampling from 2D Gaussian Mixture Model with circular obstacles.
Then, in Section~\ref{sub:mimo}, we show how adding obstacles and thereby constraining the exploration space, can accelerate sampling in a MIMO application.

\subsection{2D Gaussian mixture model with obstacles}
\label{sub:gmm}

\vspace{1mm}
\noindent\textbf{Setup.} 
We consider first a sampling task in $\ccalX = \reals^2$ where the target distribution is a mixture of two Gaussian components 
$p(\mathbf{x})=\frac{1}{2} \mathcal{N}\left(\boldsymbol{\mu}_1, \boldsymbol{\Sigma}_1\right)+\frac{1}{2} \mathcal{N}\left(\boldsymbol{\mu}_2, \boldsymbol{\Sigma}_2\right)$
with parameters 

\begin{align*}
\centering
\boldsymbol{\mu}_1 =\left[\begin{array}{l}
-2 \\
-1
\end{array}\right], 
\quad \boldsymbol{\mu}_2 =\left[\begin{array}{c}
0.9 \\
1
\end{array}\right],
\end{align*}
\begin{align*}
\centering
\boldsymbol{\Sigma}_1=\left[\begin{array}{ll}
2 & 1 \\
1 & 2
\end{array}\right], 
\quad \boldsymbol{\Sigma}_2=\left[\begin{array}{cc}
0.5 & -0.25 \\
-0.25 & 0.5
\end{array}\right].
\end{align*}
In addition, we incorporate two circular obstacles $\ccalO_i = \{\bbx \in \reals^2: (\bbx-\bbc_i)^2 - 0.4^2\}$ with $\bbc_1 = [-1, 1]^\top$ and $\bbc_2 = [-1, 0.1]^\top$; see Fig.~\ref{figs_gaussian_case} for a representation of this setting.
This setup provides a simple yet illustrative benchmark to analyze how the sampler handles multi-modal sampling and obstacle avoidance.

\vspace{1mm}
\noindent \textbf{Analysis.}
Starting from random positions $\bbx_0 \sim \ccalN(0, \bbI)$, we run Algorithm~\ref{alg:rk_sampler} for $50\times 10^4$ iterations with $\alpha = \{0.1, 1, 7\}$ and $\tau = 0.2$.
The results are shown in Fig.~\ref{figs_gaussian_case}.
As discussed in Section~\ref{subsec:rk_sampler}, the proposed sampler has different behaviors for different values of $\alpha$.
In particular, $\alpha = 1$ corresponds to the best value, achieving a good trade-off between sampling from the target distribution while avoiding the obstacles.
For $\alpha = 0.1$, it is clear that the repulsive field is too strong, pushing the particles away even from the modes of the target distribution.
On the other hand, for $\alpha = 7$ the particles collapse to the boundary of the obstacles since the second term in~\eqref{eq:rk_langevin_disc} is negligible. In contrast, the first term becomes zero at the boundaries of the obstacles.

\subsection{MIMO detection: speeding-up the sampling} 
\label{sub:mimo}

In this section, we analyze the proposed methods in the context of symbol detection in MIMO wireless systems~\cite{mimoreview1}.

\begin{figure*}[t]
    \centering
	\begin{subfigure}{.3\textwidth}
    	\centering
    	\includegraphics[width=0.7\textwidth]{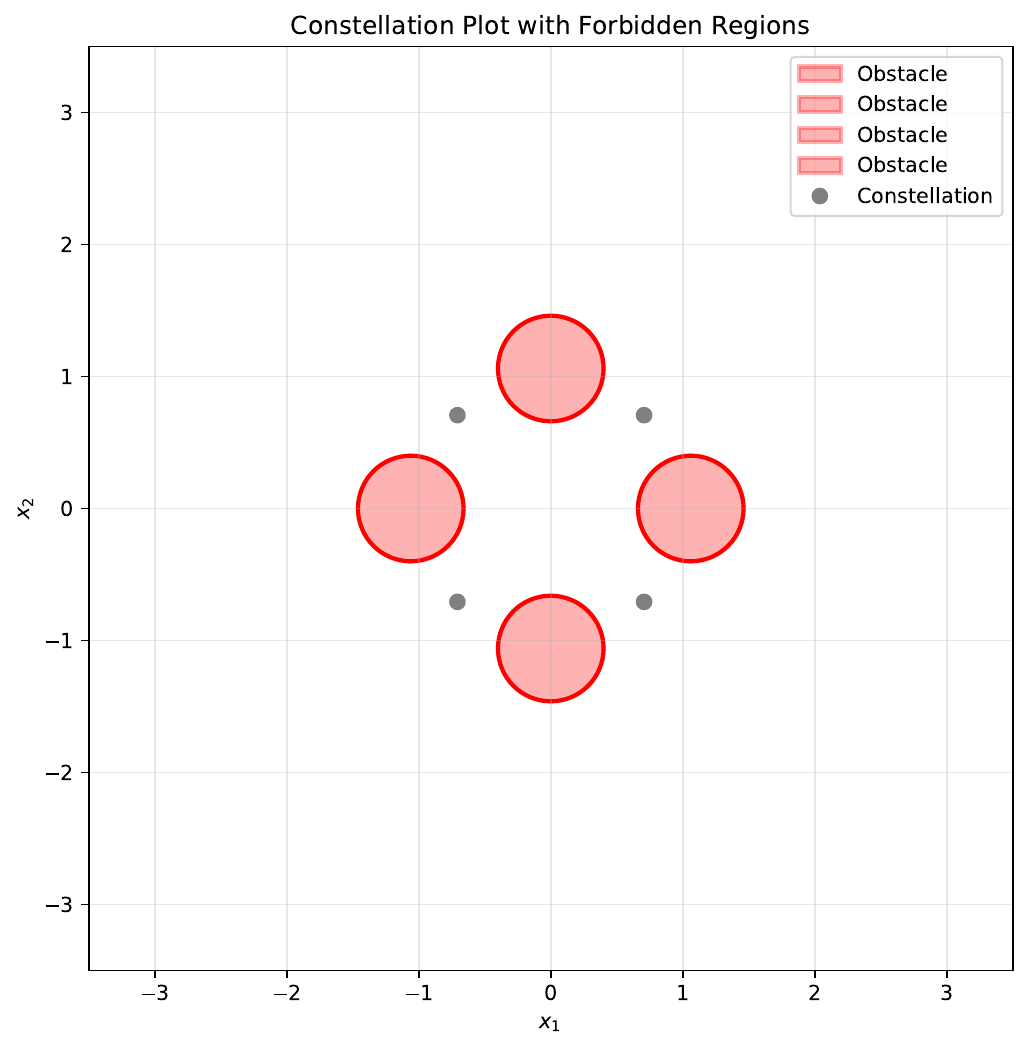}
    	\vspace{-0.1in}
    	\caption{}
    	\label{fig:mimo_scheme}
	\end{subfigure}%
	\centering
	\begin{subfigure}{.3\textwidth}
    	\centering
    	\includegraphics[width=0.7\textwidth]{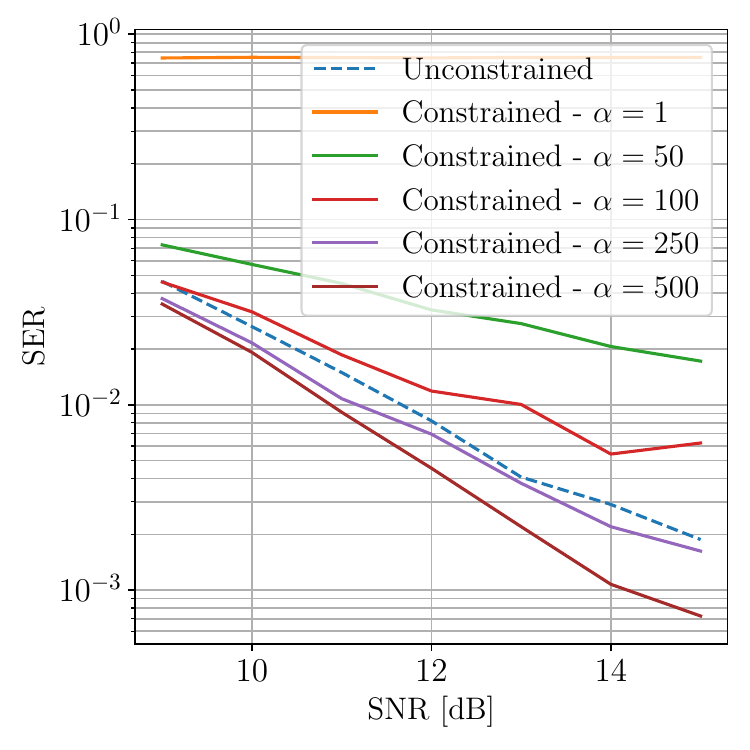}
    	\vspace{-0.1in}
    	\caption{}
    	\label{fig:ablation_mimo}
	\end{subfigure}
	\centering
	\begin{subfigure}{.3\textwidth}
    	\centering
    	\includegraphics[width=0.7\textwidth]{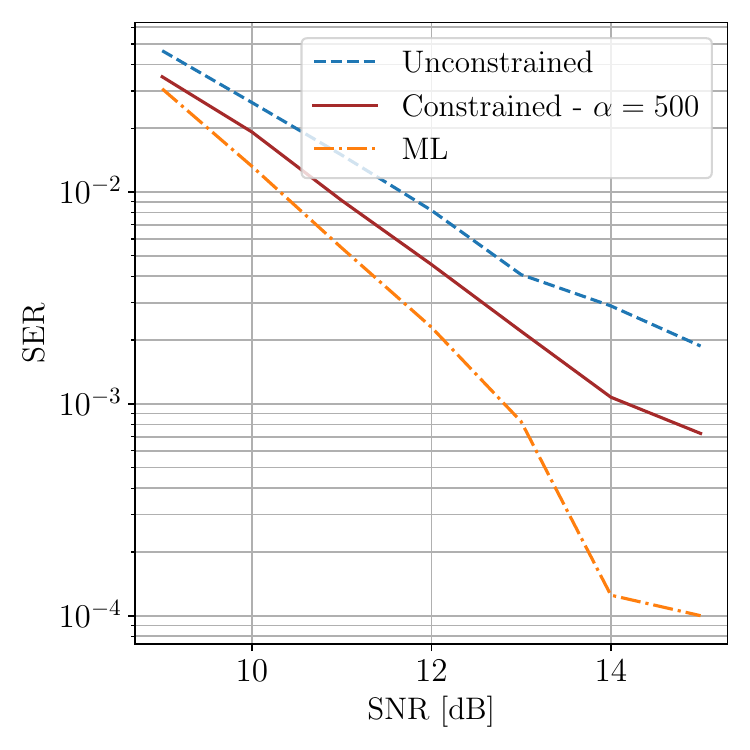}
    	\vspace{-0.1in}
    	\caption{}
    	\label{fig:baselines}
	\end{subfigure}
	\vspace{-0.025in}
	\caption{ {\small Performance analysis of our proposed methods for symbol detection estimation considering SER as a function of SNR in a Rayleigh fading channel model. 
	(a)~Constellation (QPSK) with the obstacles.
	(b)~Ablation of our proposed method with respect to $\alpha$.
    (c)~Comparison of our proposed sampler with two baselines: the unconstrained sampler and the ML, which is the optimal one.}}
	\vspace{-0.2in}
	\label{figs_channel_estimation}
\end{figure*}

\noindent \textbf{Setup and problem formulation.}
This problem can be cast as a linear inverse problem
\begin{equation}
\label{eq:inverse}
    \bby = \bbH \bbx + \bbn,
\end{equation}
where $\bbH \in \mathbb{C}^{N_r \times N_u}$ is the channel matrix, 
$\bbn \sim \mathcal{CN}(\bb0, \sigma_0^2 \bbI_{N_r})$ is complex, circular Gaussian noise,  
$\bby \in \mathbb{C}^{N_r}$ is the received vector,  
and $\bbx \in \mathcal{S}^{N_u}$ is the vector of transmitted symbols, which is \emph{unknown}.
Here, $\mathcal{S}$ denotes a finite constellation set.  
The parameters $N_u$ and $N_r$ denote the number of single-antenna users and base-station antennas, respectively.

In all experiments, we work with the real-valued representation of~\eqref{eq:inverse}.  
Given $\bby$, the goal is to estimate $\bbx$, either by computing the maximum a posteriori (MAP) estimate or by sampling from the posterior.  
Since we assume a uniform prior over constellation symbols and Gaussian measurement noise, the MAP detector coincides with the maximum-likelihood (ML) detector:
\begin{equation}\label{eq:ml}
    \vspace{-0.1cm}
	\hat{\bbx}_{\mathrm{ML}} 
	= \argmin_{\bbx \in \mathcal{X}^{N_u}} \, \|\bby - \bbH\bbx\|_2^2 .
\end{equation}
A standard approach to approximate the MAP solution is to sample $N$ candidates from the posterior and select the one minimizing~\eqref{eq:ml}.  
This strategy was proposed in~\cite{zilberstein2022}, where the detector is based on an annealed version of the Langevin dynamics in~\eqref{eq:langevin}.
The annealing acts as a continuous relaxation of the discrete symbols, progressively concentrating mass near the true discrete constellation points.  
We denote by $L$ the number of annealing levels; while the original method used $L=20$, this entails a large and mostly prohibitive running time for real-time applications.
Hence, we seek to reduce it by accelerating the sampling.

\vspace{1mm}
\noindent \textbf{Incorporating obstacles to accelerate sampling.}
While the overdamped Langevin method achieves near-optimal performance for certain channels, its runtime is too high for real-time deployment.  
Hence, there were a few attempts to accelerate the sampling process by using higher-order dynamics~\cite{zilberstein2023accelerated}.
However, they introduce auxiliary variables that increase memory usage and require additional hyperparameter tuning.  
In contrast, our approach uses the \emph{constrained} sampler to guide exploration: obstacles are introduced to repel the sampler from low-probability regions, encouraging faster traversal of the posterior landscape.  
We illustrate this idea in Fig.~\ref{fig:mimo_scheme}.

\vspace{1mm}
\noindent \textbf{Comparison with baselines.}
We consider a QPSK constellation $\mathcal{S} = \{\pm 0.74 \pm j 0.84\}$ and set $N_u = N_r = 8$.  
For the channel, we assume Rayleigh fading with $[\bbH]_{ij} \sim \mathcal{CN}(0, 1/N_r)$.  
In addition to overdamped Langevin sampling, we include as a baseline the optimal ML detector implemented using the Gurobi solver~\cite{gurobi}.
We set $L= 5$ instead of the original value $L = 20$.
We measure performance using the symbol error rate~(SER) as a function of the signal-to-noise ratio~(SNR), defined~as
\begin{equation}\label{eq:snr}
	\mathrm{SNR} = 
	\frac{\mathbb{E}[\|\bbH\bbx\|^2]}{\mathbb{E}[\|\bbn\|^2]}.
\end{equation}
For SER computation, we generate $10$ samples and retain the best candidate.  
We first perform an ablation over the parameter $\alpha$, confirming that obstacles indeed improve exploration; see Fig.~\ref{fig:ablation_mimo}.  
In particular, for $\alpha > 250$ we observe a clear improvement over the unconstrained case.  

Next, fixing $\alpha = 500$, we compare against the ML detector (Fig.~\ref{fig:baselines}).  
Although a performance gap remains relative to optimal ML, especially at high SNR, our constrained sampler significantly outperforms the unconstrained Langevin baseline for the same reduced number of iterations.

\section{Conclusion}

We introduce shielded LMC, a Langevin-based sampler designed for non-convex domains with convex obstacles. 
Inspired by navigation functions, our method modifies both the potential and diffusion terms to enforce feasibility while preserving efficient exploration of the target distribution.
Specifically, it combines a location-dependent adaptive temperature with a repulsive drift to ensure constraint satisfaction without compromising sampling diversity.
Experiments on a 2D Gaussian mixture model with convex constraints and  MIMO symbol detection demonstrate that shielded LMC reliably respects constraints, accelerates convergence, and improves downstream performance compared to unconstrained Langevin dynamics.
Future work includes establishing theoretical guarantees for the proposed dynamics, evaluating performance with more complex obstacles, and exploring integration with diffusion models for real-world applications.


\bibliographystyle{IEEEbib_abbrev}
\bibliography{citations}

\end{document}